\documentclass[twocolumn,showpacs,preprintnumbers,amsmath,amssymb]{revtex4}
\usepackage{color}
\def\mop{{\langle\mathcal{O}^H_n\rangle}}
\def\OP#1#2#3#4{{\bigl.^{#1}\hspace{-1mm}{#2}_{#3}^{[#4]}}}
\def\ME#1#2#3{{\langle\mathcal{O}^{#1}_{#2}#3\rangle}}

\def\NO{\nonumber}

\def\md{\mathrm{d}}

\def\s{\sigma}

\usepackage{graphicx,epsfig}

\def\jpsi{{J/\psi}}
\def\be{\begin{equation}}
\def\ee{\end{equation}}
\def\bea{\begin{eqnarray}}
\def\eea{\end{eqnarray}}
\def\bec{\begin{center}}
\def\eec{\end{center}}
\def\gev{\mathrm{~GeV}}

\def\upb{\Upsilon(1S)}
\def\upc{\Upsilon(2S)}
\def\upd{\Upsilon(3S)}
\def\upa{\Upsilon(1S,2S,3S)}

\begin{document}

\title{Complete next-to-leading-order study on the yield and polarization of $\upa$ at the Tevatron and LHC}

\author{Bin Gong, Lu-Ping Wan, Jian-Xiong Wang and Hong-Fei Zhang}
\affiliation{
Institute of High Energy Physics, Chinese Academy of Sciences, P.O. Box 918(4), Beijing, 100049, China.
}%
\date{\today}

\begin{abstract}
Based on nonrelativistic QCD factorization scheme, we present the first complete next-to-leading order study on the yield 
and polarization of $\Upsilon(1S,2S,3S)$ hadroproduction.
By using the color-octet long-distance matrix elements
obtained from fits of the experimental measurements on $\Upsilon$ yield and polarization at the Tevatron and LHC,
our results can explain the measurements on the yield very well,
and for the polarizations of $\Upsilon(1S,2S,3S)$,
they are in (good, good, bad) agreement with recent CMS measurement, but still have some distance from the CDF measurement.  
\end{abstract}
\maketitle

The well-known $J/\psi$ polarization puzzle became obvious when the CDF measurement at the Tevatron~\cite{Affolder:2000nn} was found 
completely different from the leading-order (LO) theoretical prediction in the framework of nonrelativistic QCD 
(NRQCD)~\cite{Beneke:1995yb}, which was proposed as a factorization approach on heavy quarkonium decay and production~\cite{Bodwin:1994jh}.
Even with the progress in the next-to-leading order (NLO) QCD calculation, theoretical studies~\cite{Butenschoen:2012px, 
Chao:2012iv,Gong:2012ug} on $J/\psi$ polarization at NLO could not clearly clarified the situation. 
Early measurements~\cite{Acosta:2001gv,Abazov:2008za}
on polarization of $\Upsilon$ at the Tevatron is in conflict with the corresponding LO NRQCD prediction~\cite{Braaten:2000gw} too. 
Recently, a very important and interesting measurement on the polarization of $\upa$ at the LHC was reported by the CMS 
collaboration~\cite{Chatrchyan:2012woa} with employing improved consideration in the measurement~\cite{Faccioli:2010ej}.
Since bottom is almost three times as heavy as charm, the NRQCD $\alpha_s$ and velocity expansions  
are of better convergence and the theoretical predictions at QCD NLO are more reliable for $\Upsilon$ 
than that for $J/\psi$. Therefore, it is very important to extend the theoretical predication on $\Upsilon$ at QCD NLO to 
solve or clarify the long-standing polarization puzzle when there are already measurement at the LHC. 

In last six years, there is some very important progress in the NLO QCD correction calculation.
The NLO corrections to color-singlet (CS) $\jpsi$ hadroproduction have been investigated in Refs.~\cite{Campbell:2007ws,Gong:2008sn},
its transverse momentum ($p_t$) distribution is found to be enhanced by $2-3$ order of magnitude at high $p_t$ region, 
and its polarization changes from transverse into longitudinal at NLO~\cite{Gong:2008sn}.
The results are reproduced at $p_t$ LO in a new factorization scheme for large $p_t$ quarkonium production~\cite{Kang:2011mg}.
The NLO corrections to $\jpsi$ production via $S$-wave color-octet (CO) states ($\OP{1}{S}{0}{8}, \OP{3}{S}{1}{8}$) are studied in Ref.~\cite{Gong:2008ft} and the corrections to $p_t$ distributions of both $\jpsi$ yield and polarization are small. In Refs.~\cite{Ma:2010vd},
NLO corrections for $\chi_{cJ}$ hadroproduction are studied.
The complete NLO calculation for prompt $\jpsi$ hadroproduction (with $\OP{3}{P}{J}{8}$ included) was given by two groups~\cite{Ma:2010yw, Butenschoen:2010rq},
and their predictions for $p_t$ distributions agree with the experimental measurements at the Tevatron and LHC.
The calculation for polarization of direct $\jpsi$ hadroproduction at NLO QCD was presented by two groups~\cite{Butenschoen:2012px, Chao:2012iv}.
The complete NLO calculation of the polarization for prompt $\jpsi$ hadroproduction was completed by our group~\cite{Gong:2012ug} last year.
It is known that at large $p_t$ region the logarithm term $\ln(p_t/m_c)$
may ruin fix order perturbative expansion and it is resummed in the new factorization scheme mentioned above~\cite{Kang:2011mg}.
But it is unclear that how large the $p_t$ region is where fix order calculation works well for $J/\psi$ case.

For $\Upsilon$ hadroproduction, there are studies on the $p_t$ distribution of yield and polarization for the CS channel at QCD NLO~\cite{Campbell:2007ws,Gong:2008sn,Gong:2008hk} and
at the partial next-to-next-to-leading order~\cite{Artoisenet:2008fc}.
NLO QCD correction to $p_t$ distribution of the 
yield and polarization for $\Upsilon(1S,3S)$ via S-wave CO states is presented in Ref.~\cite{Gong:2010bk}, and
NLO QCD correction to $p_t$ distribution of
the yield for $\Upsilon(1S)$ via all the CO states (include $^3P_J^8$) is presented in Ref.~\cite{Wang:2012is}.
The complete NLO study on polarization of $\Upsilon$ hadroproduction has not yet been achieved since there are
more complicated feeddown than charmonium case. However, the advantages for 
study on $\Upsilon$ are also obvious.  Since bottom is almost three times as heavy as charm, 
both QCD coupling constant $\alpha_s(\sqrt{4m_Q^2+p_t^2})$ and $v^2$ ($v$ is the velocity of heavy quark in the meson rest frame) 
are smaller,  and the perturbative calculation is of better convergence in the double expansion of $\alpha_s$ 
and $v^2$ on bottomonium than that on charmonium.
Furthermore, it is known that fix order calculation should be good enough at intermediate $p_t$
region although the logarithm term $\ln(p_t/m_Q)$ needs to be resummed at large $p_t$ region, hence fix order prediction 
on $\Upsilon$ hadroproduction will be very good for $p_t$ up to $60$ GeV if that on $J/\psi$ is very good for $p_t$ up to $20$ GeV, 
where $20$ GeV is a very conservative estimate.
In other words, it is expected that the theoretical predictions on the polarization and yield of $\Upsilon$ at QCD NLO should be 
in better agreement with experimental measurement up to large $p_t$ than that of charmonium.
Therefore, a full study on the polarization and yield on $\Upsilon$ at QCD NLO is
a very interesting and important task to fix the heavy quarkonium polarization puzzle while there are
already polarization measurement on $\upa$ by the CMS.
In this Letter, we present the first complete NLO study on the polarization and yield of $\upa$ based on NRQCD factorization scheme.

According to the NRQCD factorization formalism, the cross section for hadroproduction of $H$ is expressed as
\bea
d\s[pp\rightarrow H+X]&=&\sum_{i,j,n}\int dx_1dx_2 G^i_pG^j_p \NO \\
&\times&\hat{\s}[ij\rightarrow (b\bar{b})_nX]\mop,
\label{eqn:nrqcd}
\eea
where $p$ is either a proton or anti-proton, the indices $i, j$ run over all
the partonic species and $n$ denotes the color, spin and angular momentum states of the intermediate $b\bar{b}$ pair. It can be $\OP{3}{S}{1}{1}$,$\OP{3}{S}{1}{8}$,$\OP{1}{S}{0}{8}$ and $\OP{3}{P}{J}{8}$ for $\Upsilon$, or $\OP{3}{P}{J}{1}$ and $\OP{3}{S}{1}{8}$ for $\chi_{bJ}$.
The short-distance contribution $\hat{\sigma}$ can be calculated perturbatively, while the long-distance matrix elements (LDMEs) $\mop$ are fully governed by non-perturbative QCD effects.

The polarization of $\Upsilon$ is described by three parameters, as defined in Ref.~\cite{Beneke:1998re}:
\be
\lambda=\frac{\md\sigma_{11}-\md\sigma_{00}}{\md\sigma_{11}+\md\sigma_{00}},\mu=\frac{\sqrt{2}\mathrm{Re}\md\sigma_{10}}{\md\sigma_{11}+\md\sigma_{00}},\nu=\frac{2\md\sigma_{1,-1}}{\md\sigma_{11}+\md\sigma_{00}},\NO
\ee
where $d\sigma_{S_zS_z^\prime}$ is the spin density matrix of $\Upsilon$ hadroproduction.
In this work, we focus on polarization parameter $\lambda$ in helicity frame only.

To obtain $\md\sigma_{S_zS_z^\prime}$, similar treatment as in Ref.~\cite{Gong:2012ug} is taken for both direct and
feeddown contributions. There are various feeddown contributions in $\Upsilon$ production, while some of them are ignored in our calculation as they are thought to be small. The feeddown contributions included in this work are:
\begin{itemize}
\item $\Upsilon(3S)$: no feeddown contribution is included.
\item $\Upsilon(2S)$: feeddown contributions from $\Upsilon(3S)$ and $\chi_{bJ}(2P)$ are included.
\item $\Upsilon(1S)$: feeddown contributions from $\Upsilon(2S,3S)$ and $\chi_{bJ}(1P,2P)$ are included.
\end{itemize}

%
Newly updated Feynman Diagram Calculation package~\cite{Wang:2004du} is used in our calculation.

\begin{table*}[ht]
\begin{tabular}{|c|c|c|c|c|c|c|c|c|c|c|c|c|c|c|c|}
\hline
$H$  & $\upb$  & $\upc$  &  $\upd$  &  $\chi_{b0}(1P)$ & $\chi_{b1}(1P)$  & $\chi_{b2}(1P)$ & $\chi_{b0}(2P)$  &  $\chi_{b1}(2P)$  &  $\chi_{b2}(2P)$ \ \\
\hline
${\cal B}(H\rightarrow\mu\mu)(\%)$    &2.48     & 1.93    & 2.18     &   $-$            &      $-$         &      $-$        &        $-$       &          $-$      &  $-$  \\
\hline
${\cal B}[H\rightarrow\upb](\%)$      &  $-$      &  26.5   & 6.6      &   1.76              &  33.9              &  19.1             &  0.9             &  10.8            &  8.1              \\
\hline
${\cal B}[H\rightarrow\upc](\%)$      & $-$     &  $-$    & 10.6     &   $-$            &      $-$         &     $-$         &  4.6             &  19.9              & 10.6   \\
\hline
$M_H$(GeV) &  9.5    &  10.023 &  10.355  &  9.859           &  9.893           &  9.912          &  10.23           &  10.255           &  10.269             \\
\hline
\end{tabular}
\caption{Branching ratios and masses of bottomonia are taken from PDG~\cite{Beringer:1900zz}. }
\label{table:paras}
\end{table*}
In our numerical calculation, the CTEQ6M PDFs~\cite{Pumplin:2002vw} and corresponding two-loop
QCD coupling constant $\alpha_s$ are used. Branching ratios and masses of involving bottomonia can be found in 
Tab.~\ref{table:paras}. Mass of bottom quark is set $m_b=M_H/2$ as an approximation, while $M_H$ is the mass of bottomonium $H$. The
CS LDMEs are estimated by using a potential model result~\cite{Eichten:1995ch}, which gives $|R_{\upa}(0)|^2=6.477, 3.234, 2.474 \gev^3$, and  $|R^\prime_{\chi_{b}(1P,2P)}(0)|^2=1.417, 1.653\gev^5$, respectively.
The renormalization and factorization scales are chosen as $\mu_r=\mu_f=m_T$, with $m_T=\sqrt{(2m_b)^2+p_t^2}$, while the NRQCD scale is chosen as $\mu_\Lambda=m_b v \approx 1.5\gev$.
The center-of-mass energy is 1.8 and 1.96 TeV for Tevatron run I and run II, and 7 TeV for the LHC, respectively.
Various rapidity cuts are chosen according to various experimental data, including both central
and forward rapidity region. Besides, a shift $p_t^H \approx p_t^{H^\prime}\times(M_H/M_{H^\prime})$
is used while considering the kinematics effect in the feed-down from higher excited states.

In Ref.~\cite{Wang:2012is}, feeddown contributions from $\OP{3}{S}{1}{8}$ channel of $\chi_{bJ}$ are included in corresponding CO LDMEs of $\Upsilon$. But when studying the polarization, we have to separate $\OP{3}{S}{1}{8}$ channels of $\Upsilon$ and $\chi_{bJ}$, as they have different behavior in polarization. Unfortunately, there is still no experimental data
for $\chi_{bJ}$ hadroproduction till now, so there is no direct clue to determine the CO LDMEs of $\chi_{bJ}$
production. Thus we take them as extra variables in our fit, and include experimental data for $\Upsilon$ polarization
as well.  It is known that the double expansion in $\alpha_s$ and $v^2$ is not good enough in the small $p_t$ regions.
Therefore, only data in the region $p_t > 8 \gev$ are used in our fit.


\begin{table}[ht]
\begin{tabular}{|c|c|c|c|}
\hline
H  &$\ME{H}{}{(\OP{1}{S}{0}{8})}$  &\ $\ME{H}{}{(\OP{3}{S}{1}{8})}$&\ $\ME{H}{}{(\OP{3}{P}{0}{8})}/m_b^2$ \\
\hline
$\upb$        & $11.15 \pm 0.43$ \   & \ $-0.41 \pm 0.24$ \   & \ $-0.67\pm0.00$ \ \\
\hline
$\upc$        & $3.55\pm 2.12$ \  & \ $0.30 \pm 0.78$ \  & \ $-0.56\pm0.48$ \ \\
\hline
$\upd$        & $-1.07 \pm 1.07$ \  & \ $2.71 \pm 0.13$ \   & \ $0.39\pm 0.23$ \ \\
\hline
$\chi_{b0}(2P)$  & $-$             \  & \ $2.76\pm0.67$ \   & \ $-$ \ \\
\hline
$\chi_{b0}(1P)$  & $-$             \  & \ $1.27\pm0.16$ \     & \ $-$ \ \\
\hline
\end{tabular}
\caption{The obtained CO LDMEs for bottomonia production (in unit of $10^{-2}\gev^3$).}
\label{table:ldmes}
\end{table}

\begin{figure*}
\begin{center}
\includegraphics*[scale=0.27]{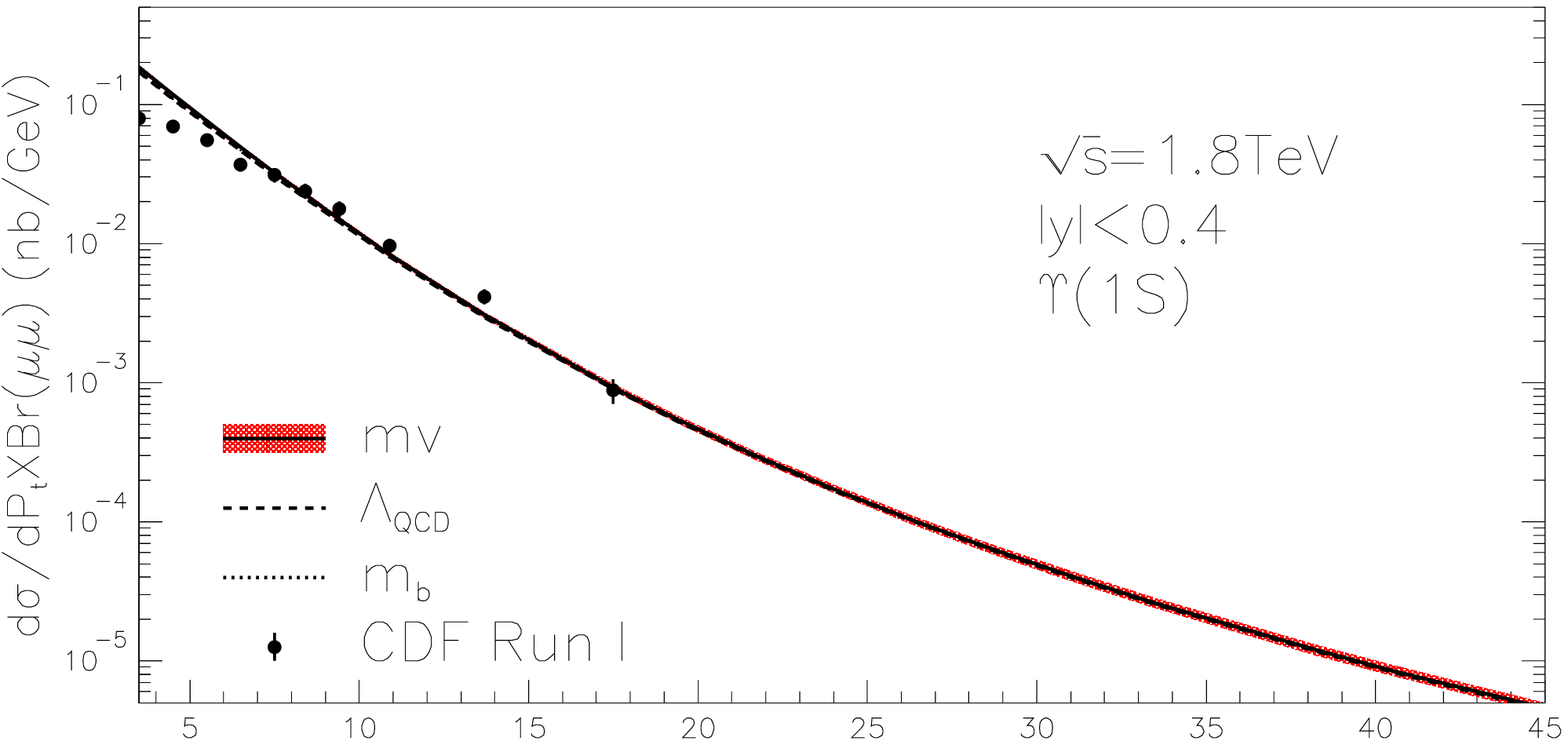}
\includegraphics*[scale=0.27]{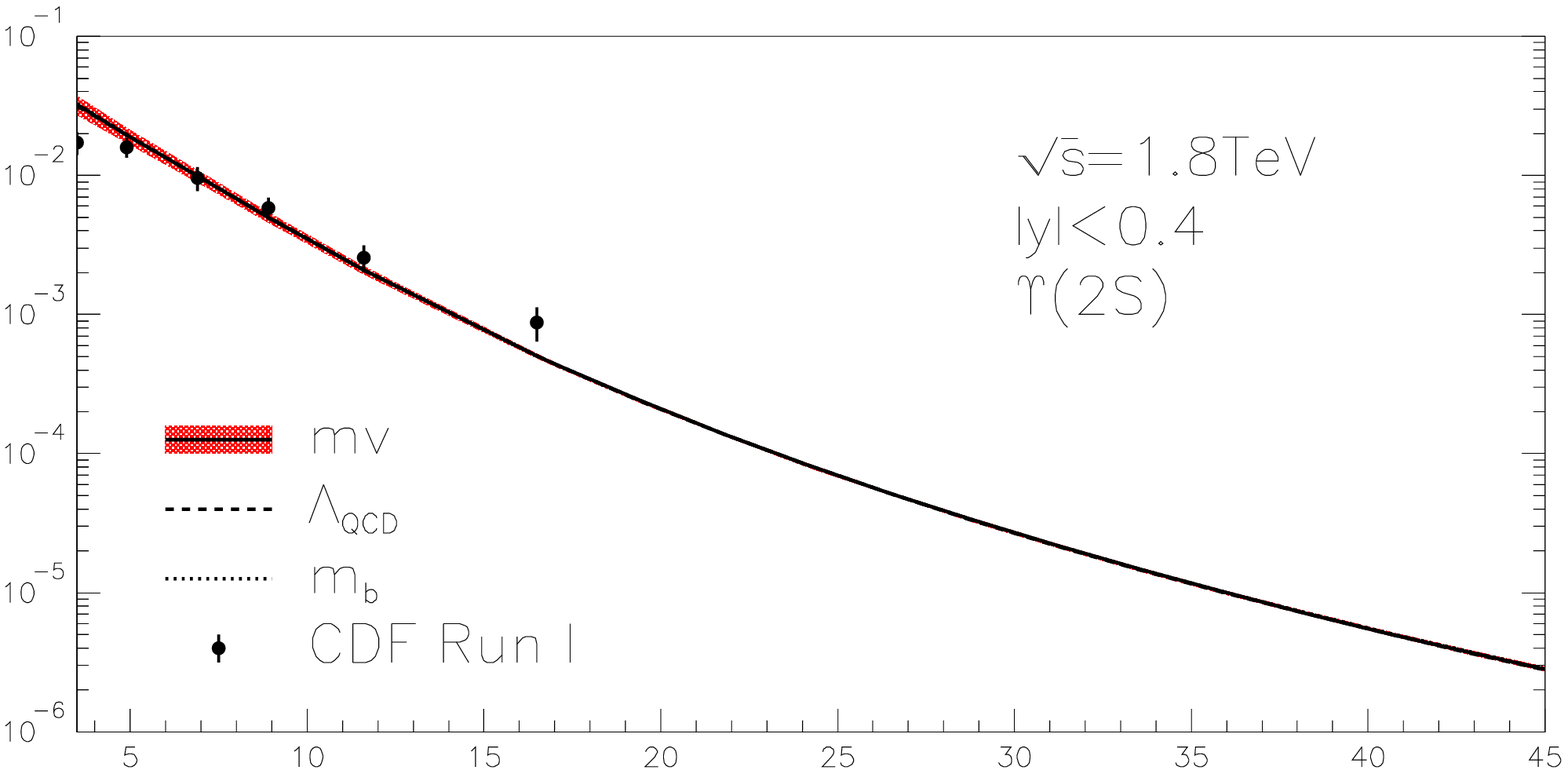}
\includegraphics*[scale=0.27]{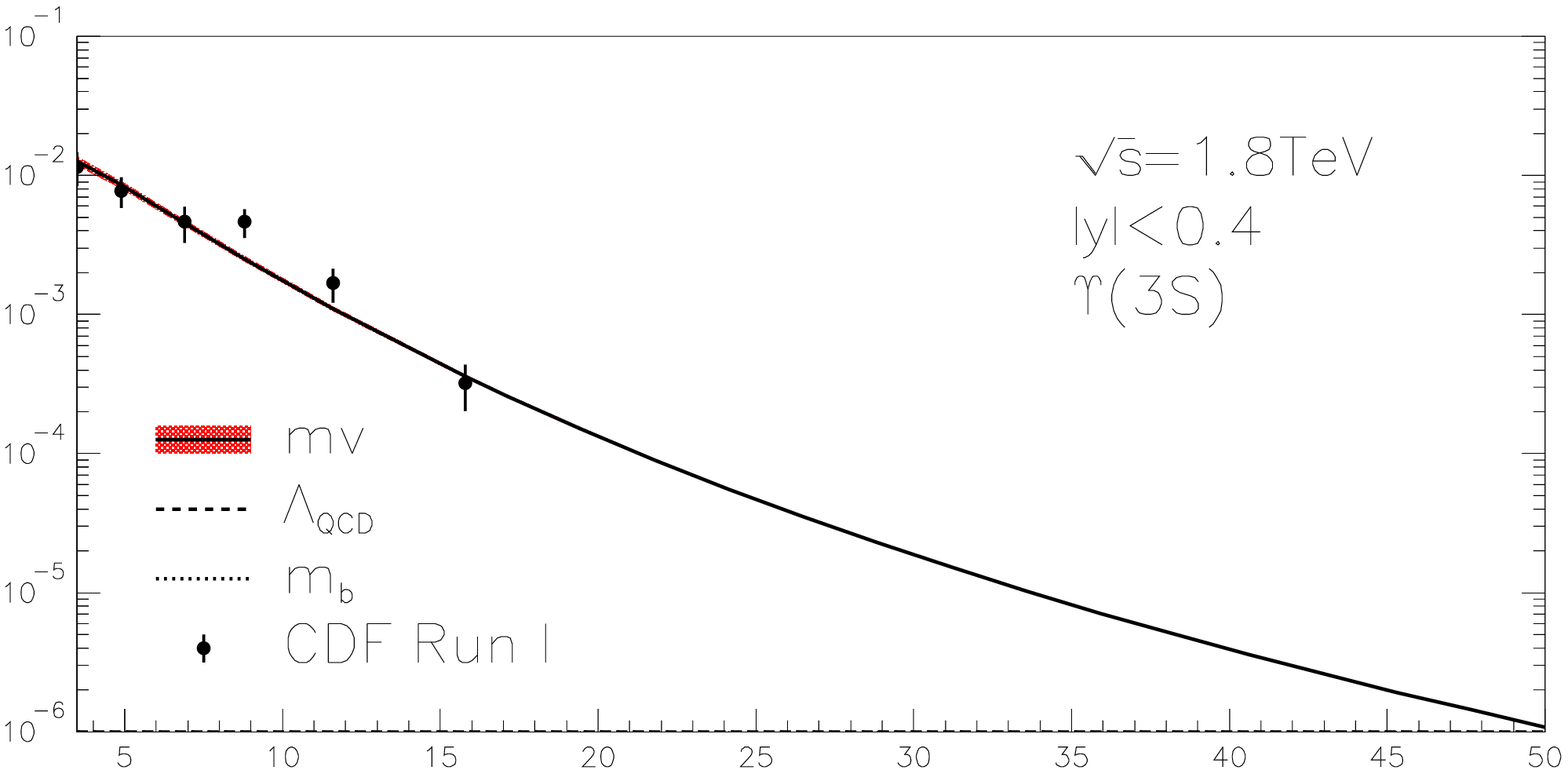}
\\
\includegraphics*[scale=0.27]{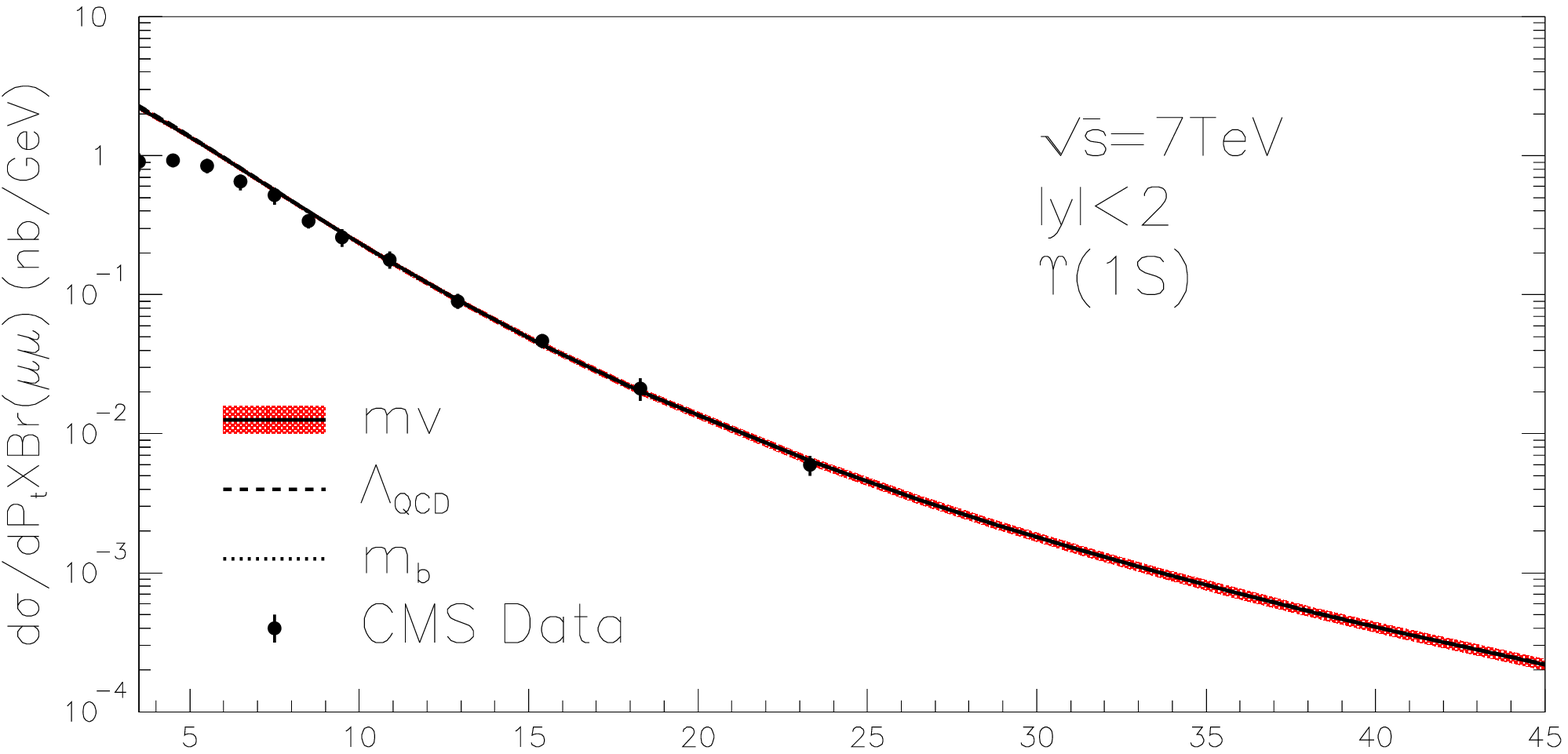}
\includegraphics*[scale=0.27]{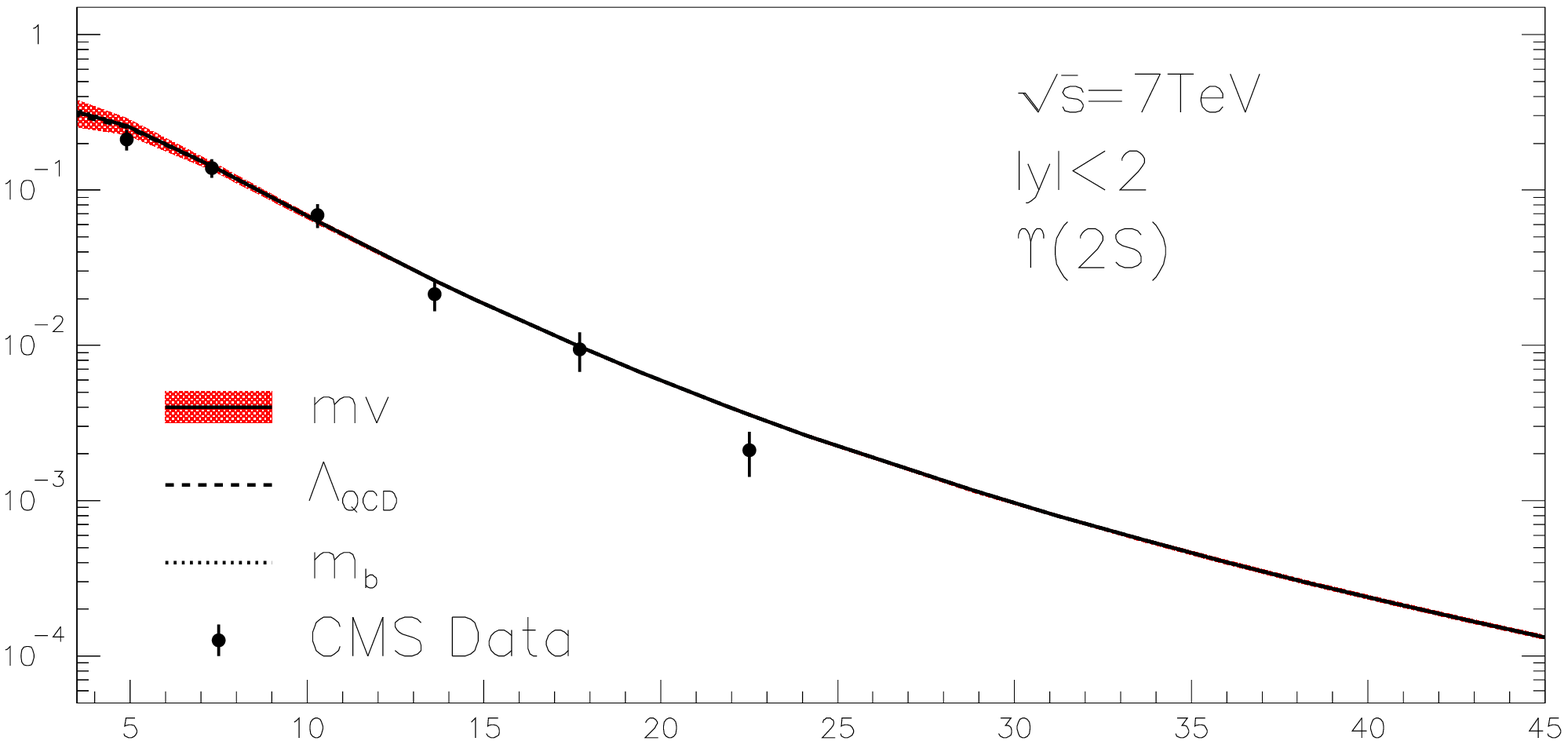}
\includegraphics*[scale=0.27]{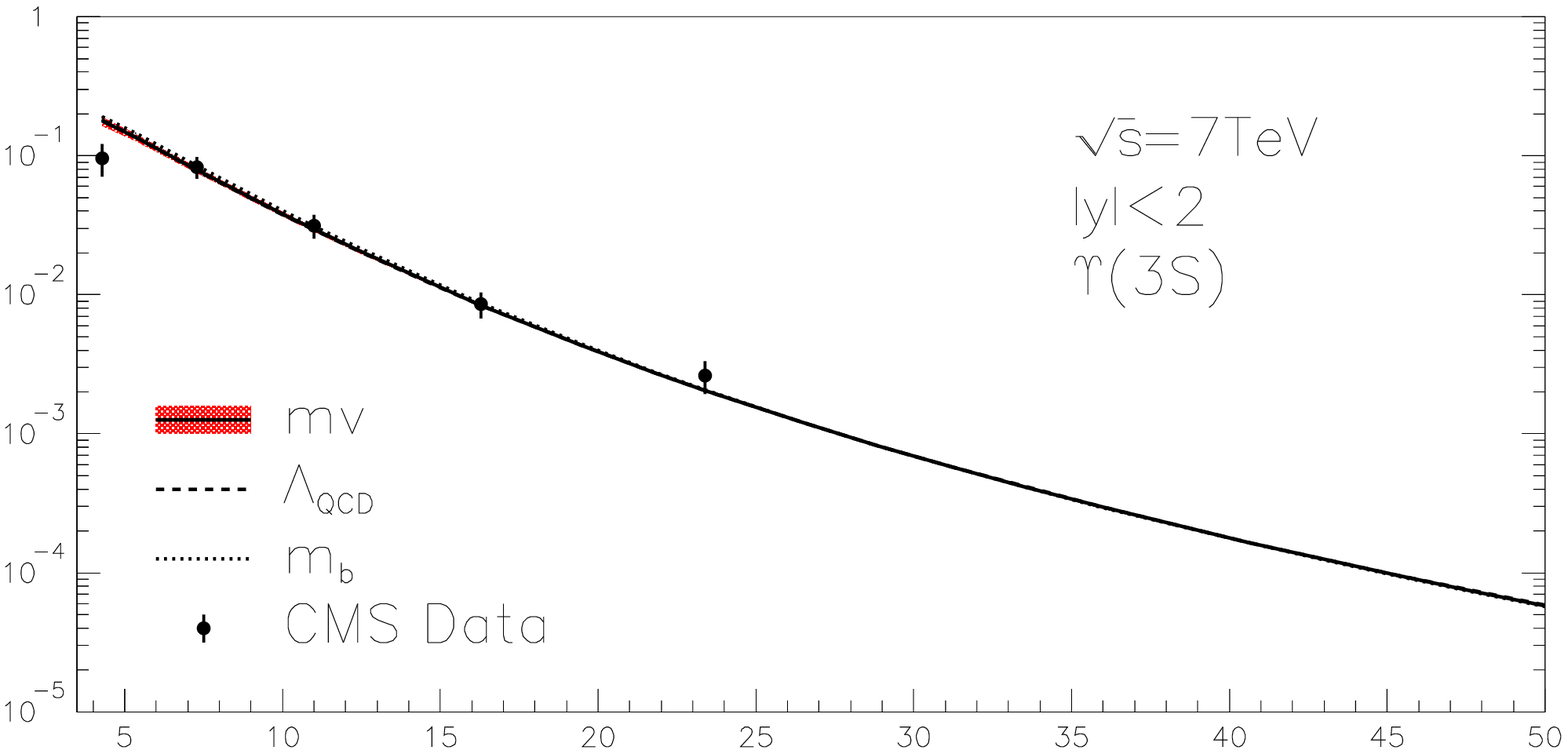}
\\
\includegraphics*[scale=0.27]{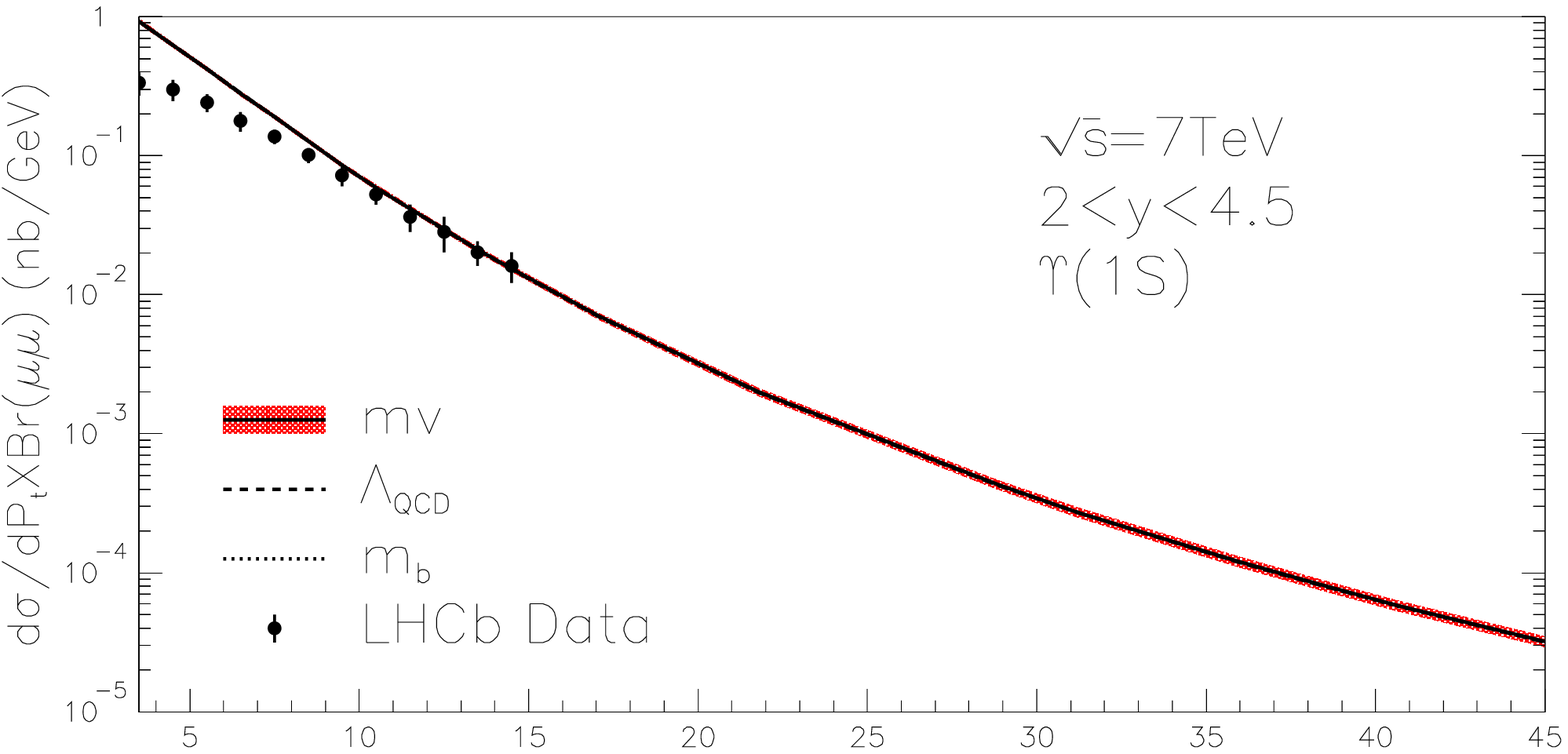}
\includegraphics*[scale=0.27]{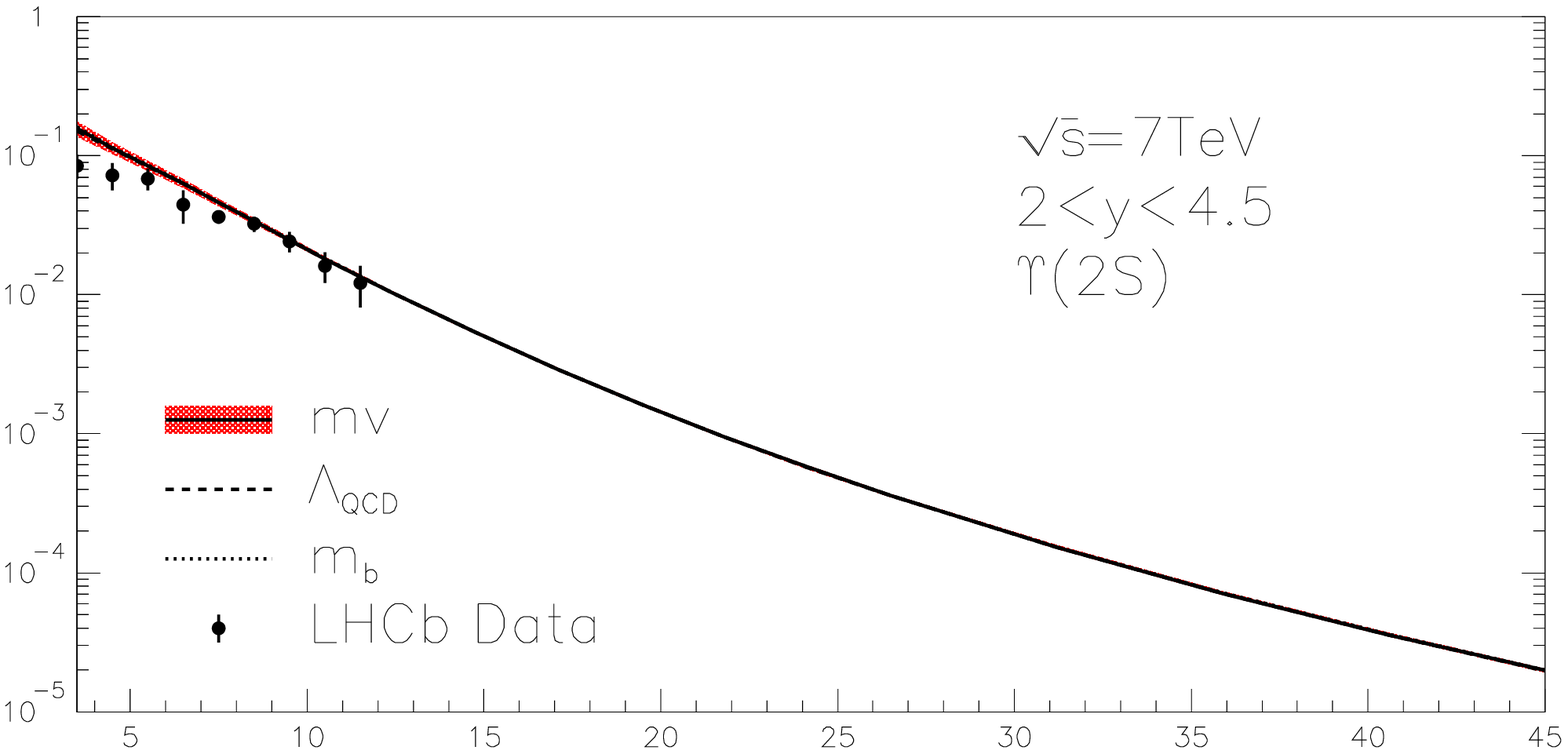}
\includegraphics*[scale=0.27]{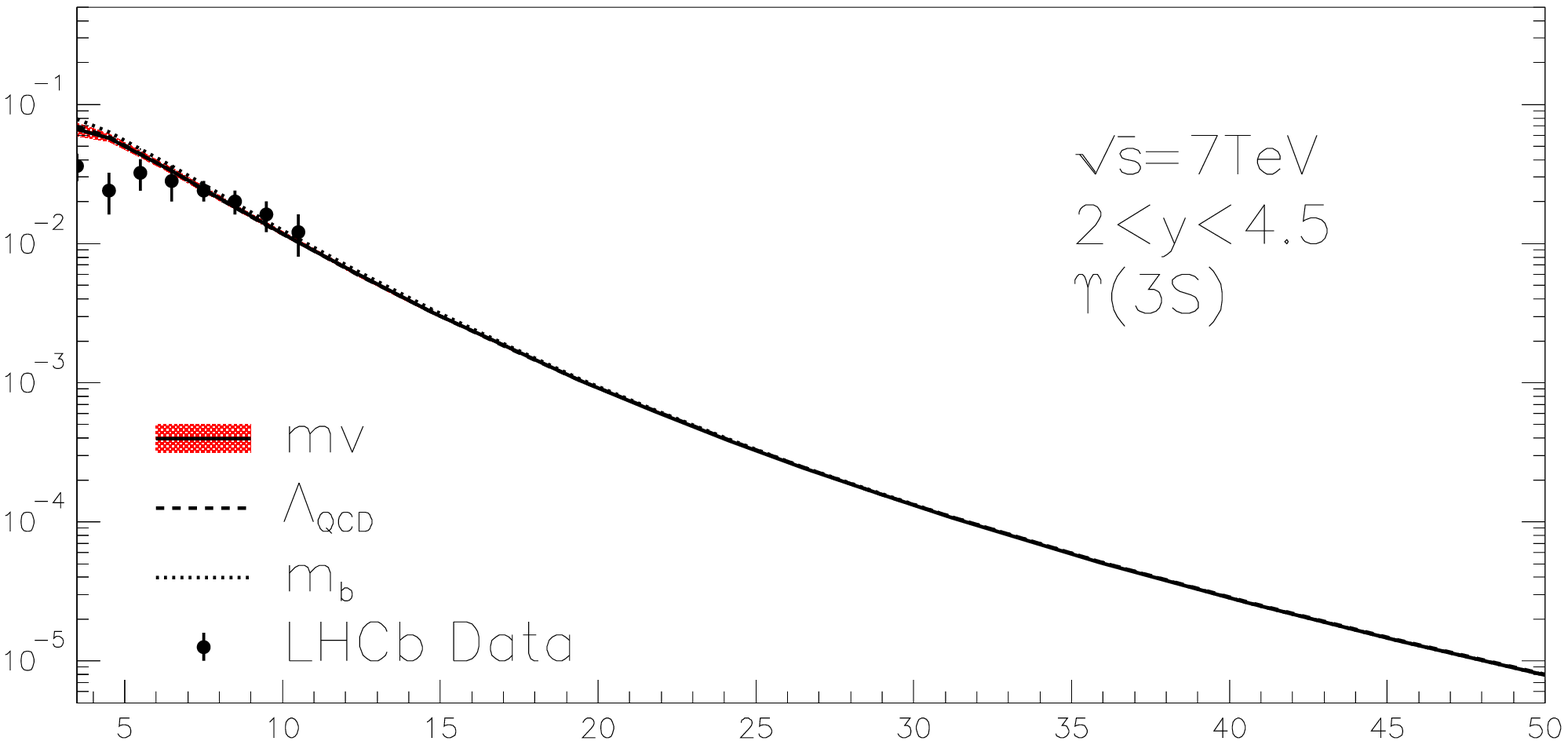}
\\
\includegraphics*[scale=0.27]{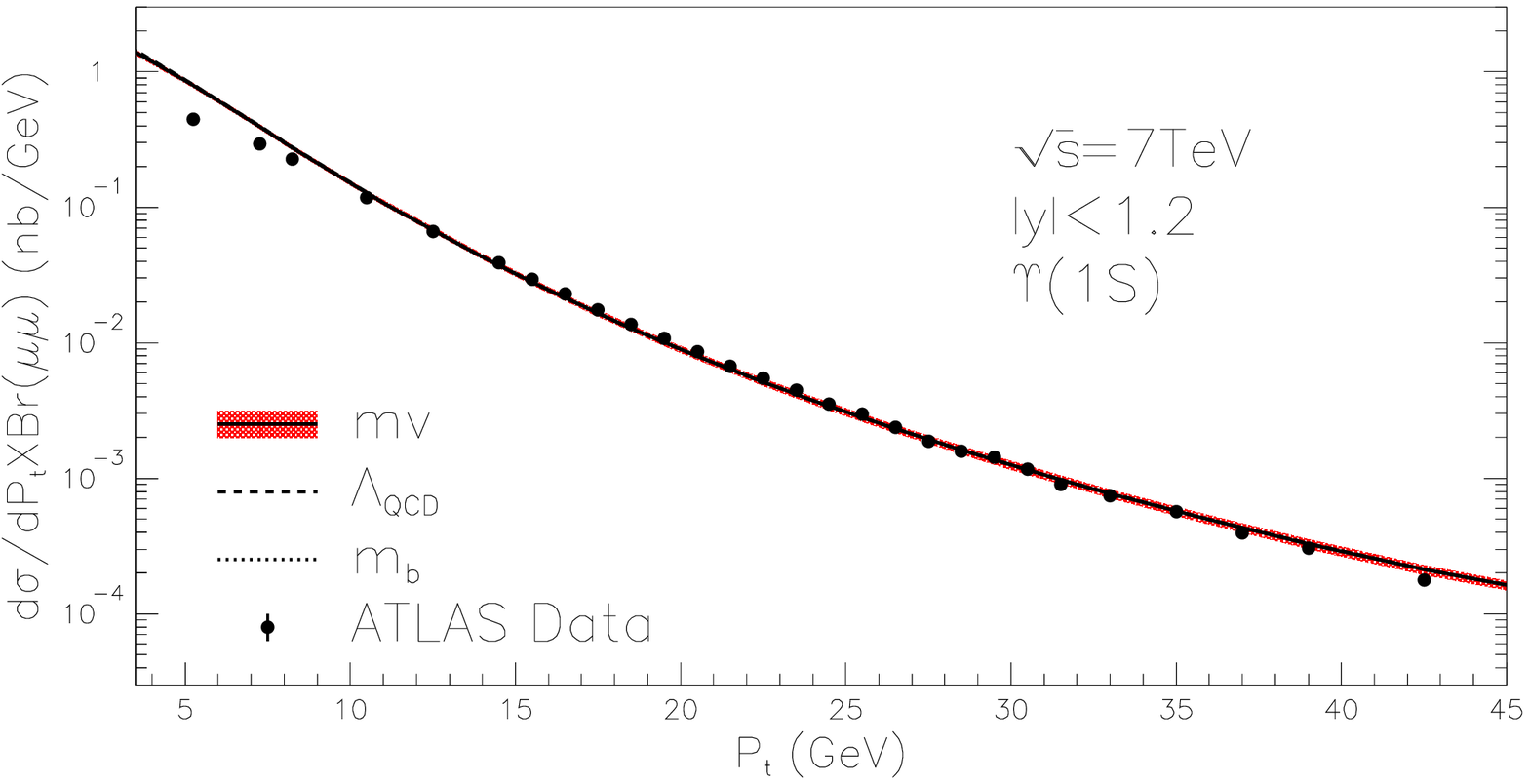}
\includegraphics*[scale=0.27]{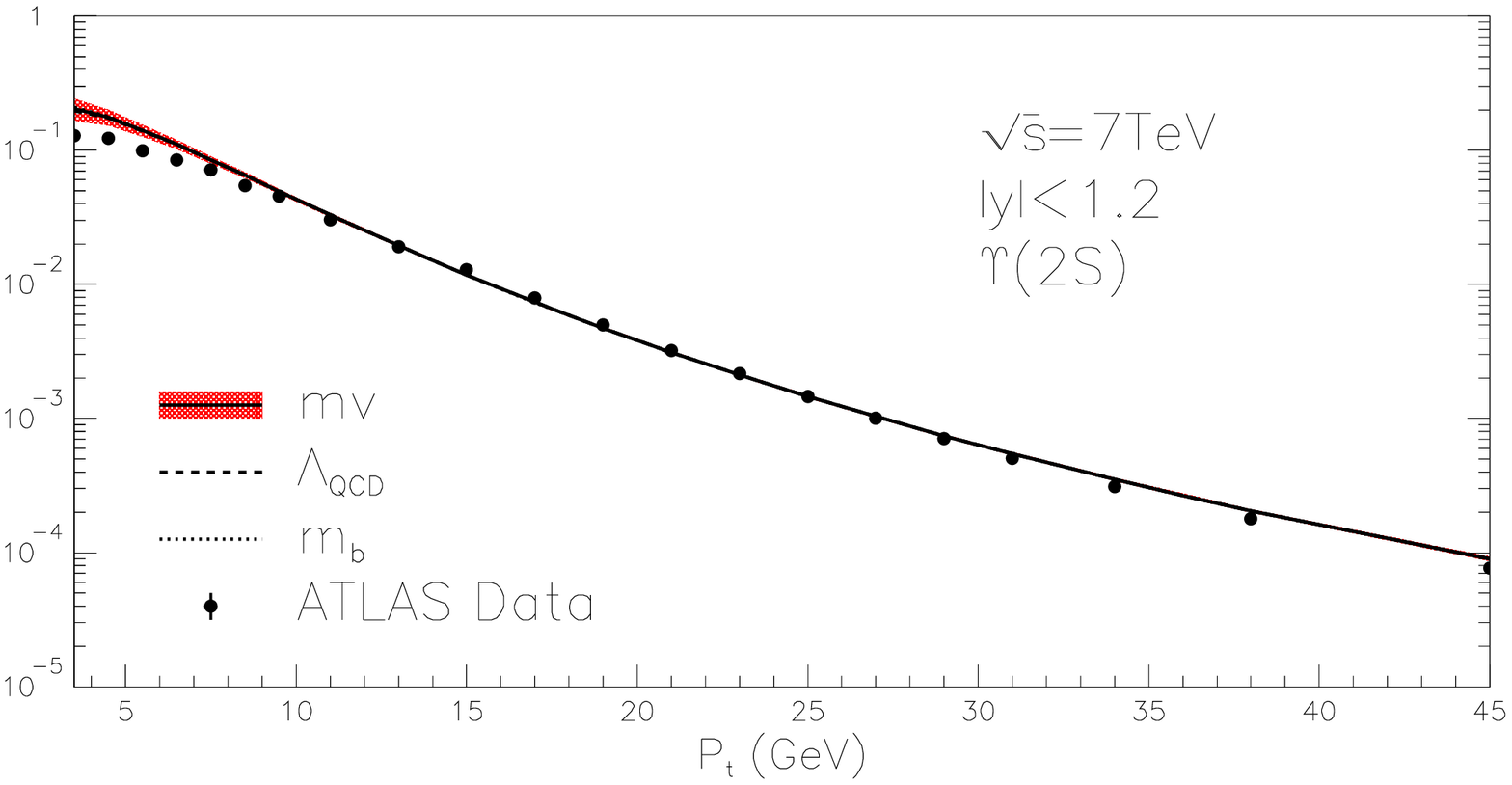}
\includegraphics*[scale=0.27]{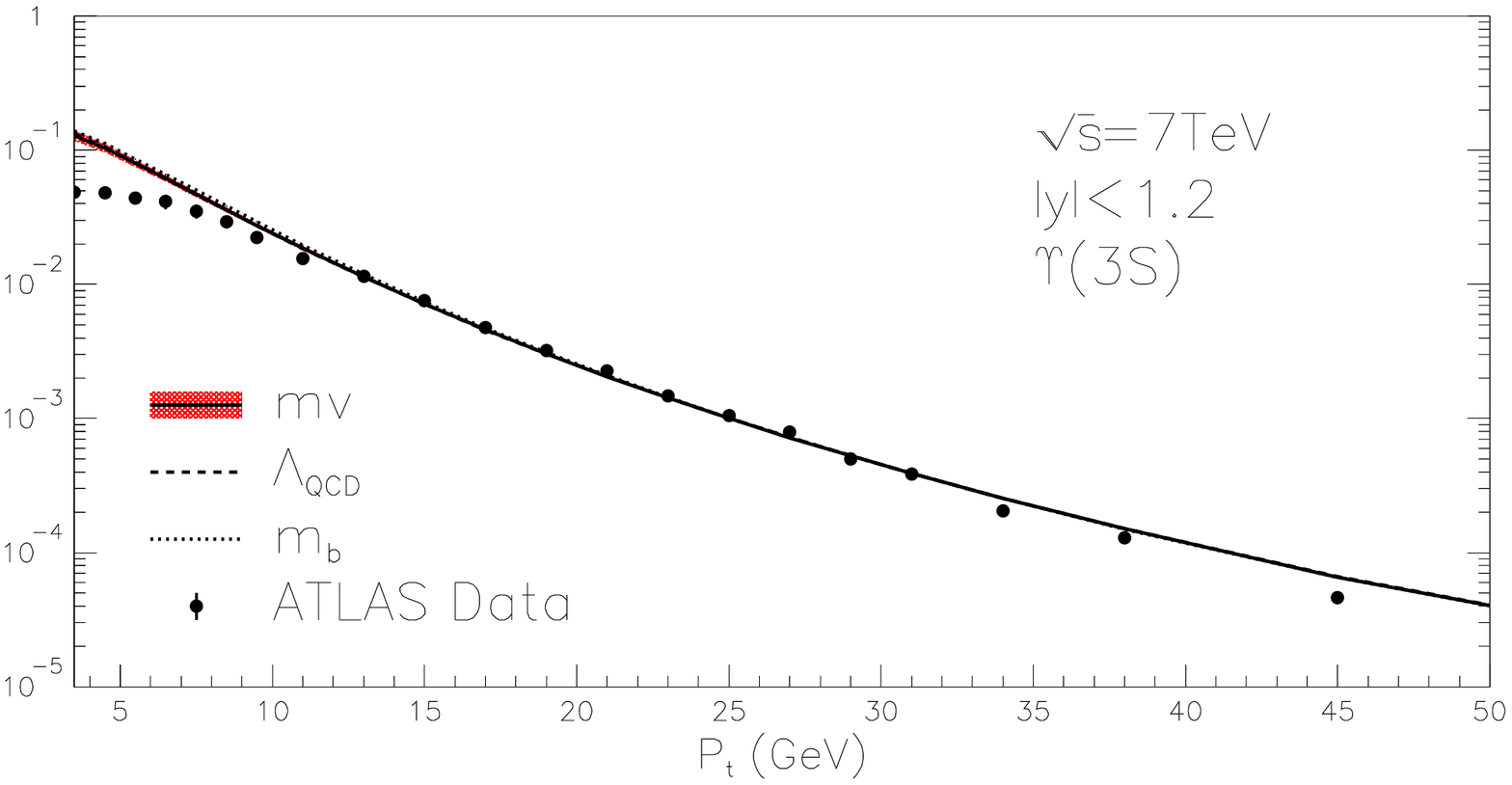}
 \caption{Differential cross section for $\Upsilon$ hadroproduction at the Tevatron and LHC. From left to right: $\Upsilon(1S)$, $\Upsilon(2S)$ and $\Upsilon(3S)$. Rows from top to bottom are corresponding to different experimental conditions of CDF RUN I, CMS, LHCb and ATLAS. The experimental data are taken from Refs.~\cite{Acosta:2001gv, LHCb:2012aa, Khachatryan:2010zg, Aad:2012dlq}.}
 \label{fig:pt}
 \end{center}
\end{figure*}

\begin{figure*}
\begin{center}
\includegraphics*[scale=0.27]{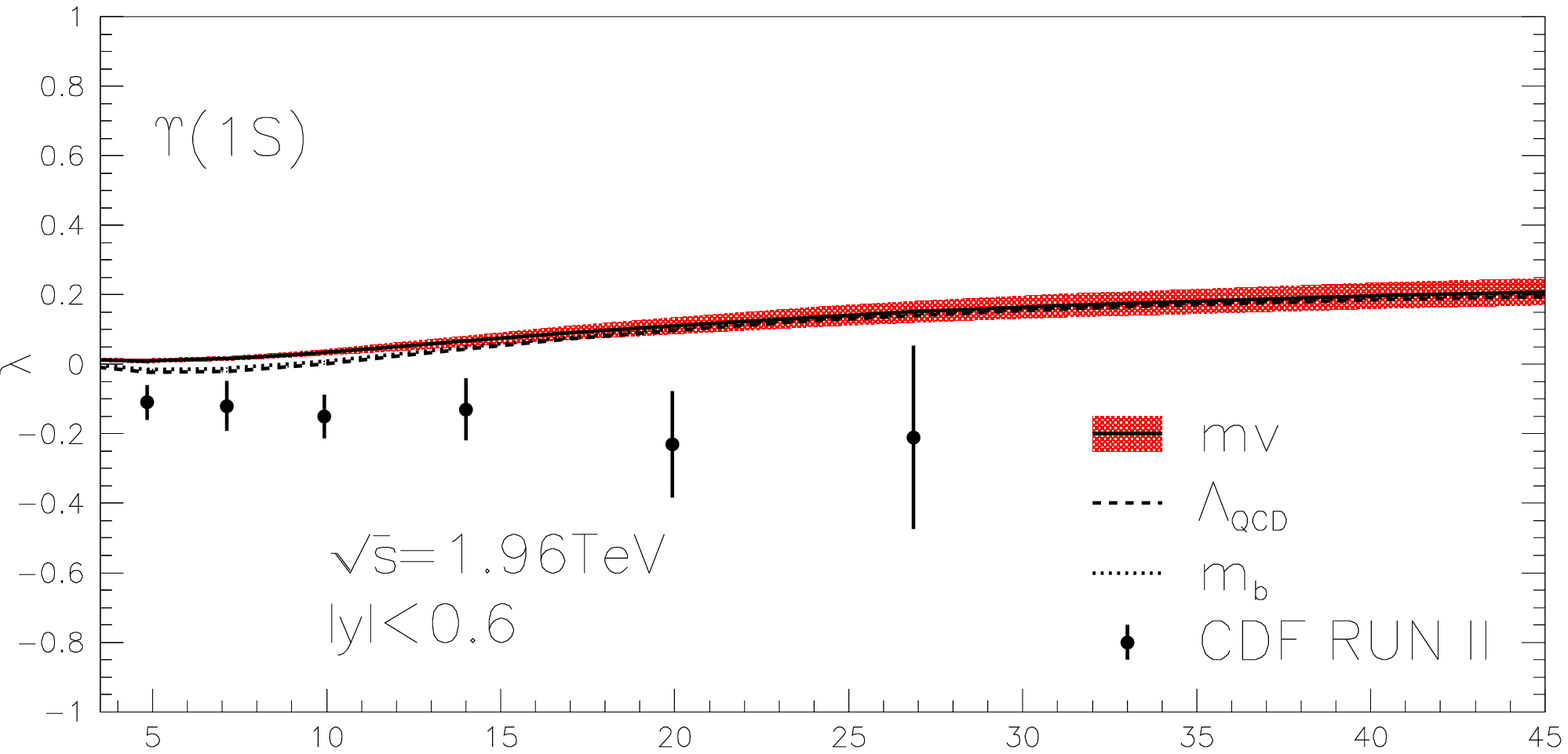}
\includegraphics*[scale=0.27]{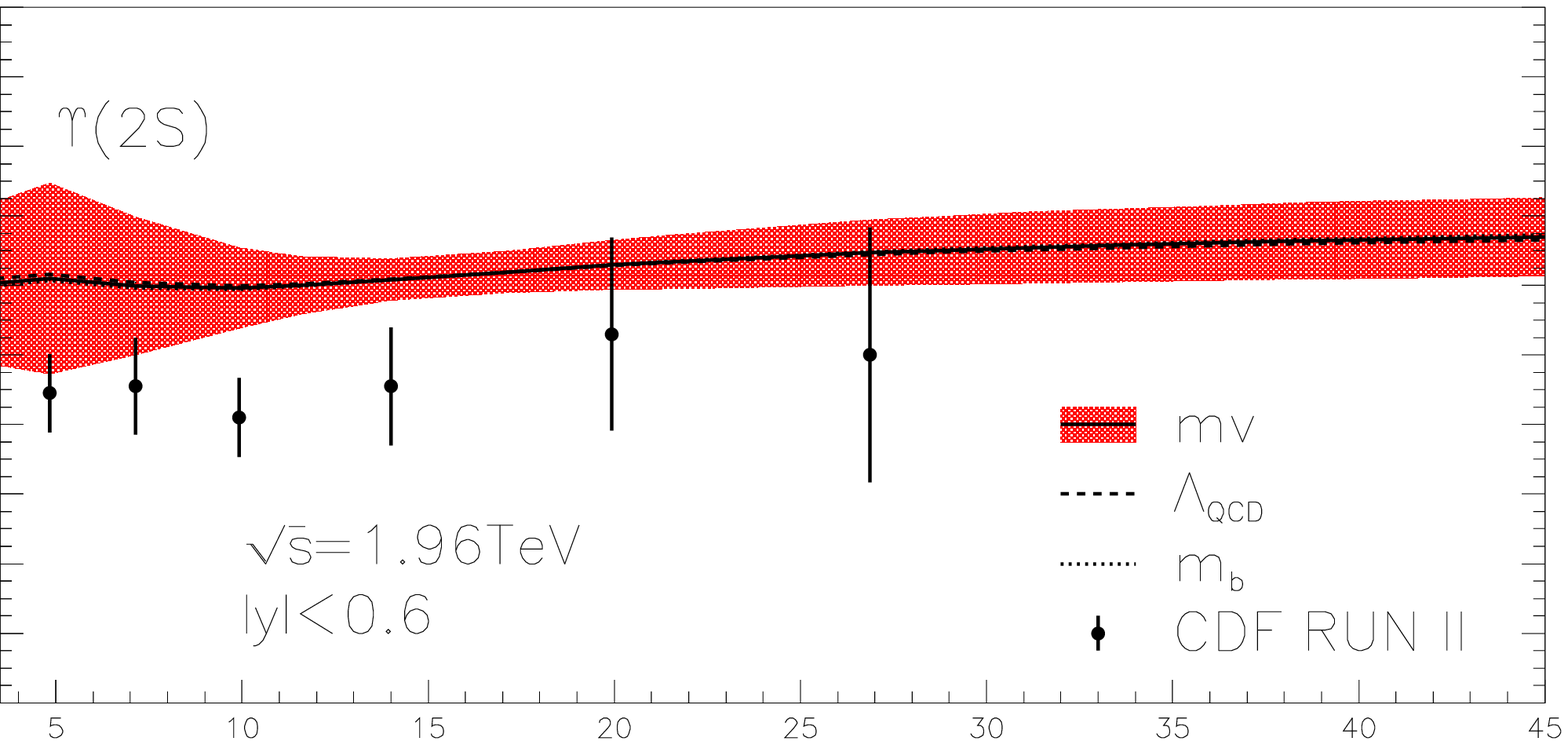}
\includegraphics*[scale=0.27]{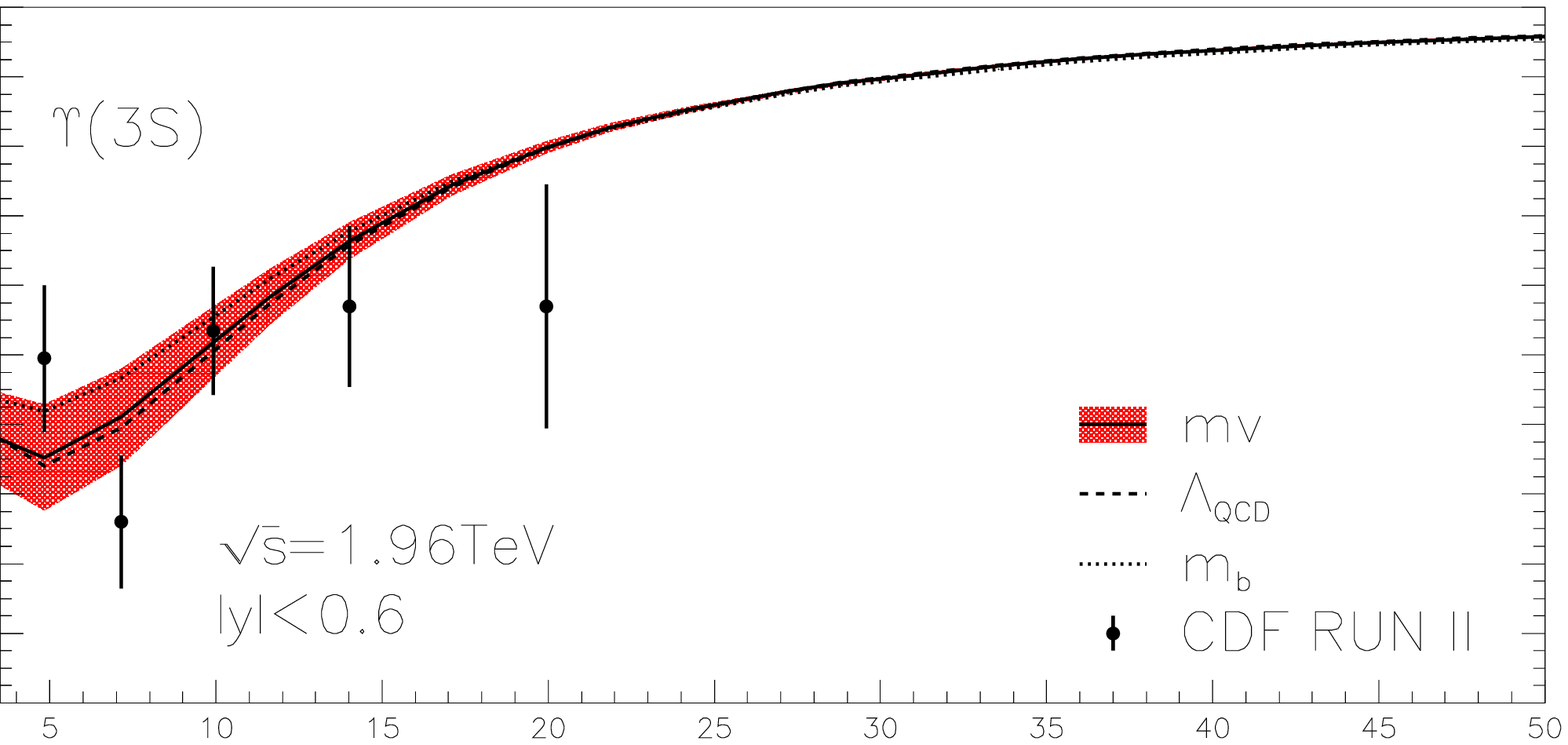}
\\
\includegraphics*[scale=0.27]{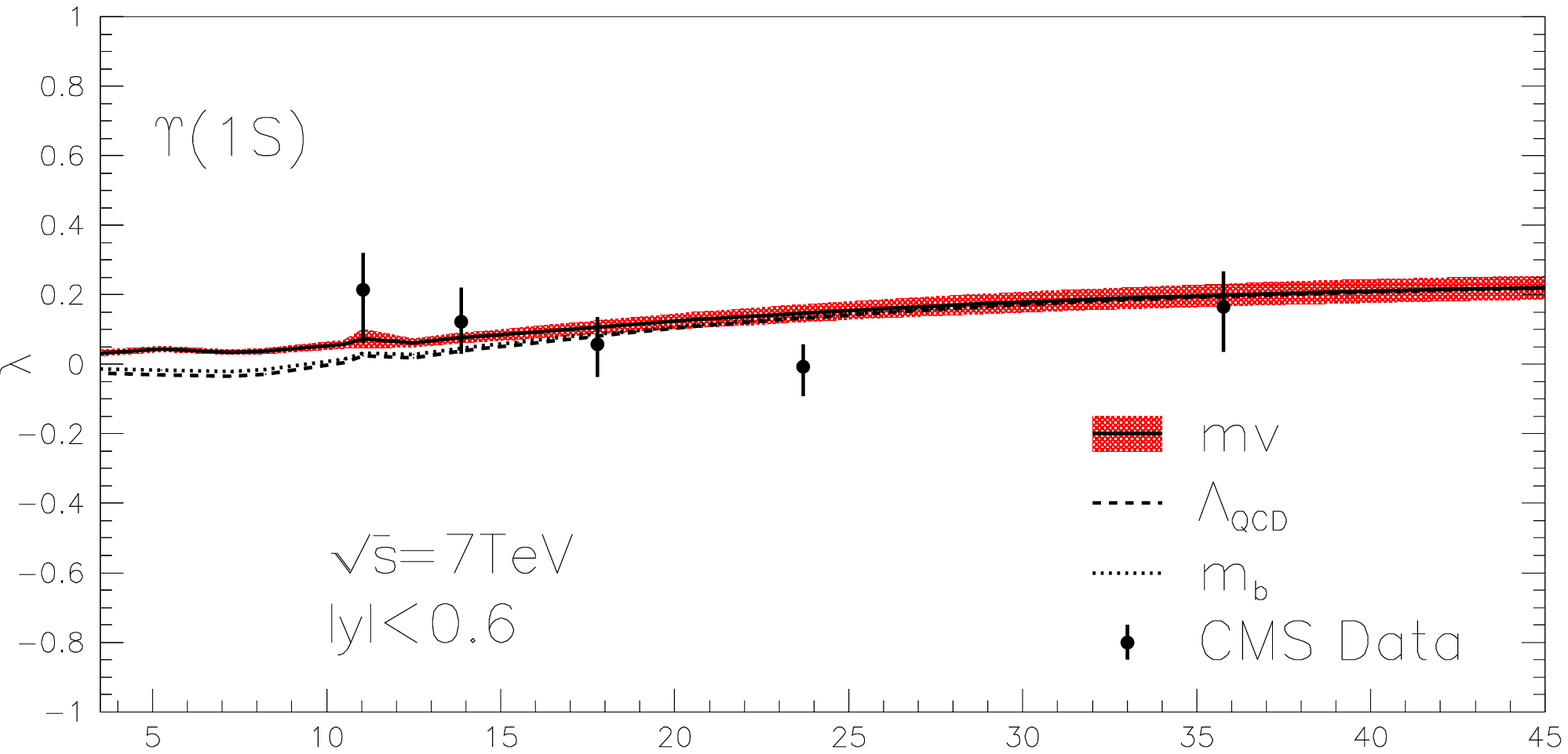}
\includegraphics*[scale=0.27]{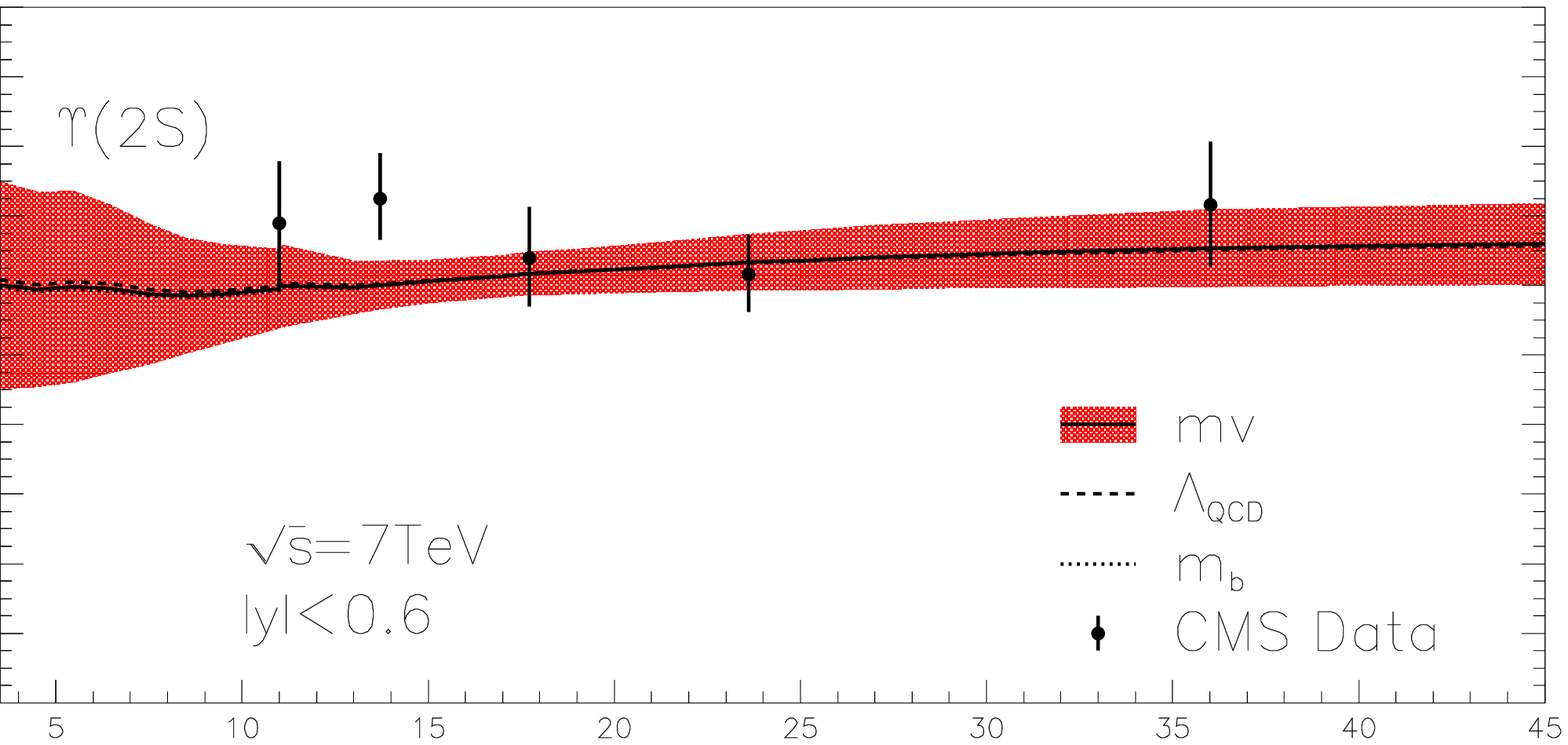}
\includegraphics*[scale=0.27]{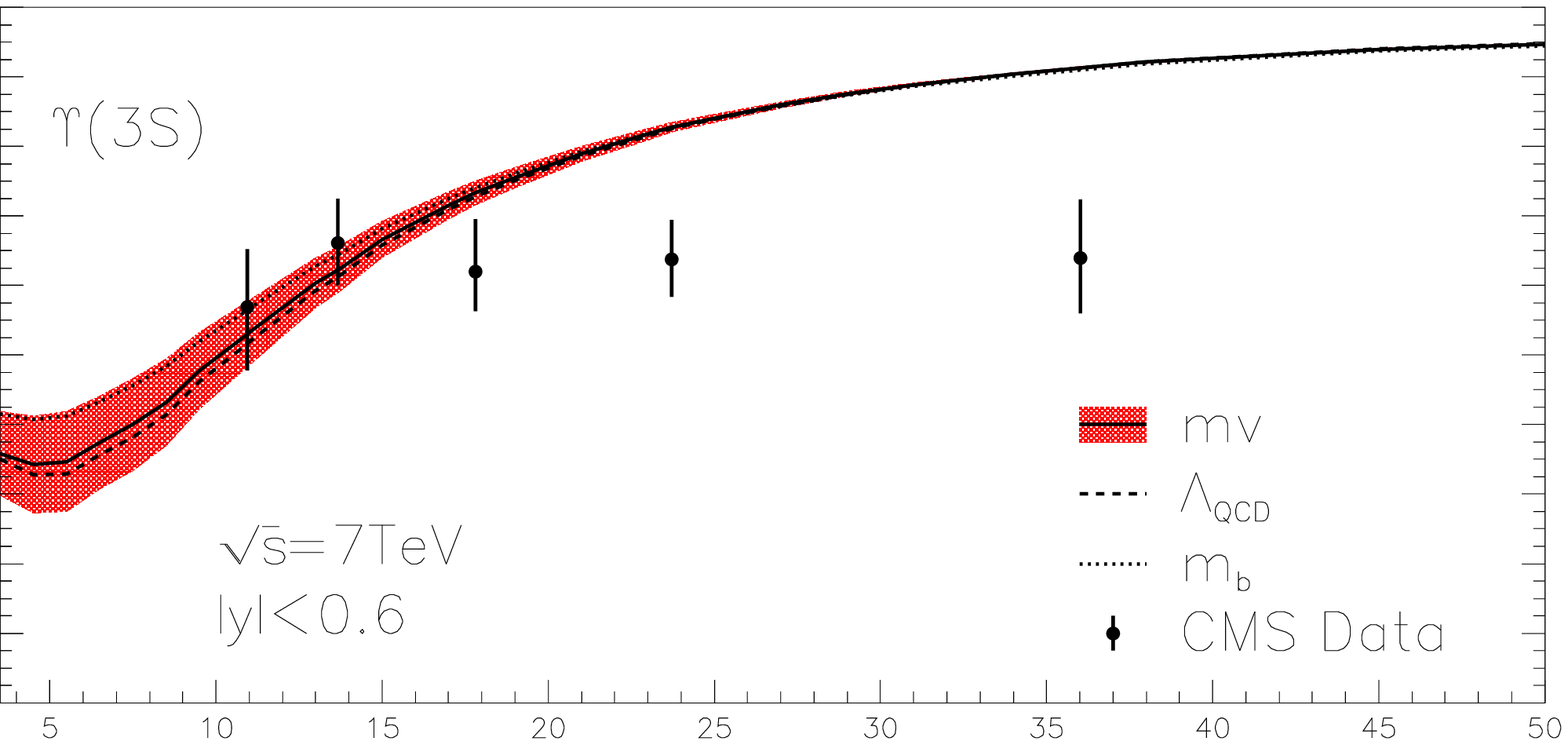}
\\
\includegraphics*[scale=0.27]{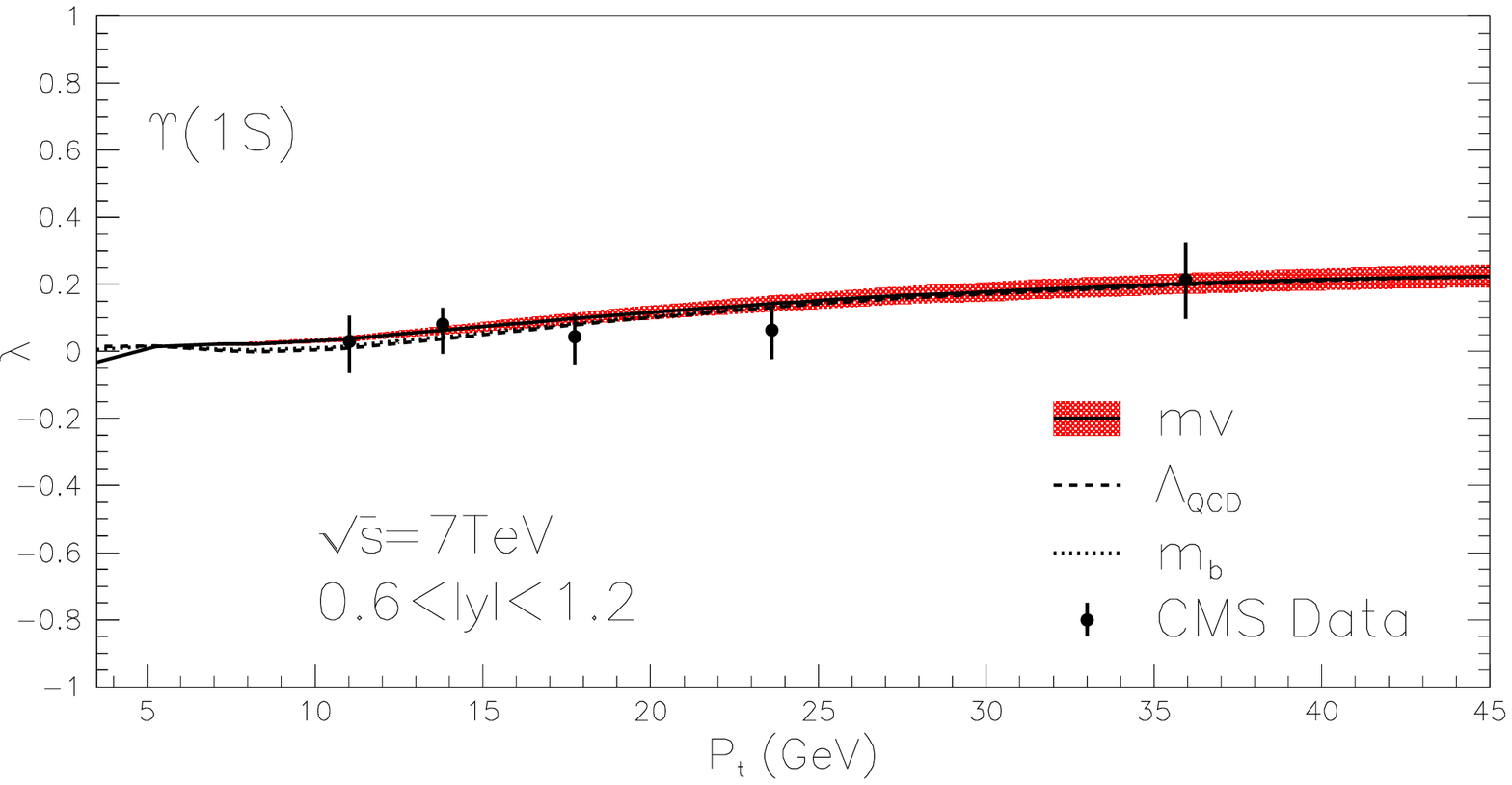}
\includegraphics*[scale=0.27]{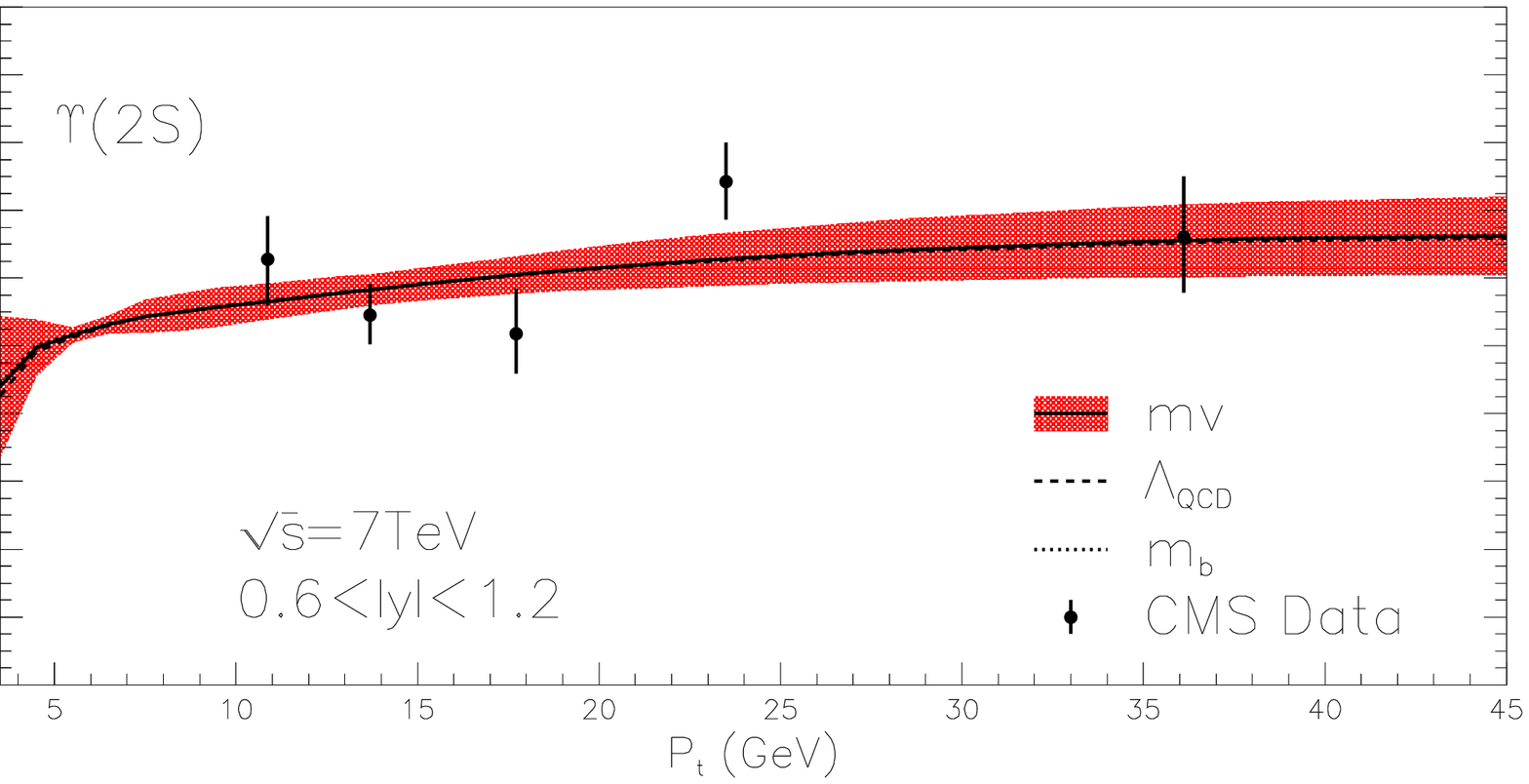}
\includegraphics*[scale=0.27]{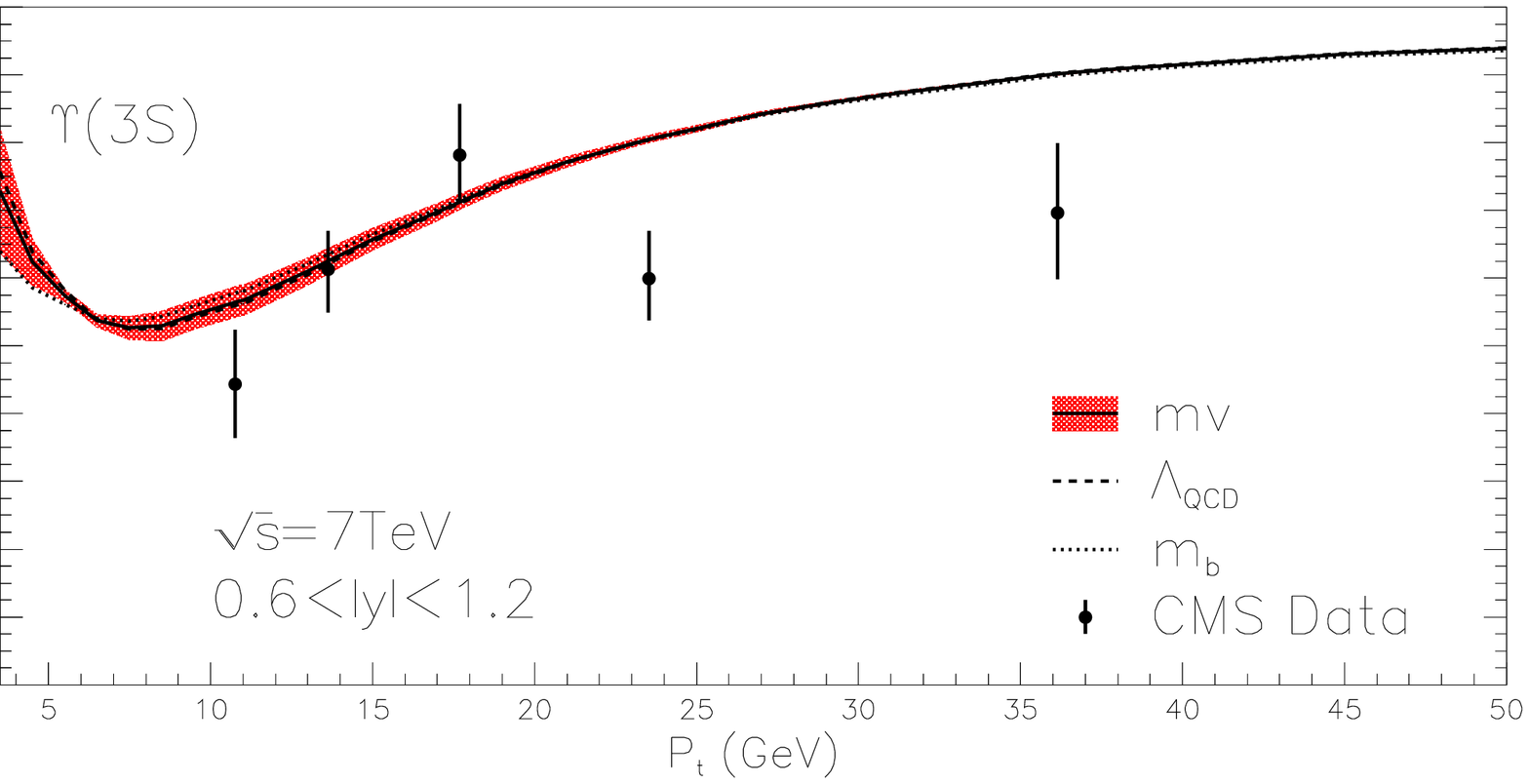}
  \caption{Polarization parameter $\lambda$ for $\Upsilon$ hadroproduction at the Tevatron and LHC. From left to right: 
$\Upsilon(1S)$, $\Upsilon(2S)$ and $\Upsilon(3S)$. Rows from top to bottom are corresponding to different experimental 
conditions of CDF RUN II, CMS ($|y|<0.6$) and CMS ($0.6<|y|<1.2$). The CMS and CDF data are taken from 
Refs.~\cite{CDF:2011ag,Chatrchyan:2012woa}.}
  \label{fig:polar}
  \end{center}
\end{figure*}

In the fit, we have used the experimental data for $p_t$ distribution of the differential cross section by 
CDF~\cite{Acosta:2001gv}, LHCb~\cite{LHCb:2012aa}, CMS~\cite{Khachatryan:2010zg} and ATLAS~\cite{Aad:2012dlq}, 
and of the polarization by CDF~\cite{CDF:2011ag} ( where early measurements
from CDF~\cite{Acosta:2001gv}  and D0~\cite{Abazov:2008za}  conflict
with CDF new measurement~\cite{CDF:2011ag} and are given up ) and CMS~\cite{Chatrchyan:2012woa}.
Three fits are performed for $\Upsilon(3S,2S,1S)$ hadroproduction step by step. 
In order to express the uncertainty from the CO LDMEs in theoretical predictions correctly, 
a covariance-matrix method is performed as Ref.~\cite{Gong:2012ug}. 
Here are some notes:
\begin{itemize}
\item $\chi^2/d.o.f.=117/37$ is obtained in the fit of $\Upsilon(3S)$.
\item  
$\chi^2/d.o.f.=88/37$ is obtained in the fit of $\Upsilon(2S)$ with four CO LDMEs, one of which is  
$\ME{\chi_{b0}(2P)}{}{(\OP{3}{S}{1}{8})}$. 
$\chi_{bJ}(2P)$ feeddown contributes a fraction of about $35\sim 76\%$ as $p_t$ increases in $\Upsilon(2S)$ hadroproduction.

\item In the fit of $\Upsilon(1S)$, we have also included data for the fraction of $\Upsilon(1S)$ from 
$\chi_{bJ}(1P)$ feeddown by LHCb~\cite{Aaij:2012se} and obtained the four CO LDMEs with $\chi^2/d.o.f.=107/63$.
Furthermore,  the obtained $\chi_{b}(1P,2P)$ fraction in $\Upsilon(1S)$ is consistent with the CDF measurement~\cite{Affolder:1999wm}.
\end{itemize}

All the fitted CO LDMEs can be found in Tab.~\ref{table:ldmes}. With these LDMEs, our predictions for
the differential cross section of $\Upsilon$ hadroproduction are shown in Fig.~\ref{fig:pt}, while those for
the polarization are shown in Fig.~\ref{fig:polar}. The uncertainty bands in the figures come from errors of the LDMEs. 
Our calculation show that it is about $25\sim 10$ ($35 \sim 20$) percent uncertainties for the differentail cross section with the 
factorization (renormalization) scale changing as $0.5\sim 2$ $m_T$, and it decreses as $p_t$ increasing. And it is smaller for 
the polarization distribution.  We do not include these uncertainties in our fitting and final plots 
since it need much more computer source consumption.

From the figures we see that the predictions on the yield of $\Upsilon$ hadroproduction can explain the experimental data
very well with very small uncertainty in a wide range of $p_t$ at the LHC and Tevatron, while for the polarization, 
things are quite different.
For $\Upsilon(3S)$, the production is dominated by $\OP{3}{S}{1}{8}$ channel, which results a transverse polarization in high $p_t$ region
and make the theoretical predictions far and far away from the experimental data as $p_t$ increases. And it is obvious that 
the polarization can not be explained at LO in $v^2$ and NLO in $\alpha_s$ if unknown feeddown contribution from higher 
excited bottomonia is negligible.
For $\Upsilon(1S,2S)$, the predictions for polarizations can explain CMS data well, but still have some distance from the measurement by CDF. 
From the measurement at the LHC, it is easy to see that the $p_t$ distribution for $\Upsilon(1S)$ is of steepest slope, and that for $\Upsilon(2S)$ is of steeper slope than that for $\Upsilon(3S)$.
For the CO contribution, we can see that $p_t$ distribution of $^1S_0^8$ is the steepest one and that of $^3S_1^8$ is the most flat one while that of $^3P_J^8$ is sensitive to the choice of NRQCD factorization scale $\mu_\Lambda$. 
From the numerical results, we obtained that the $\chi_{bJ}$ feeddown contribution in $\Upsilon(1S)$ becomes dominant as $p_t$ increases, and polarization via this channel is slightly transverse polarized,
combined with the fact that the direct part is dominated by $^1S_0^8$ channel at small $p_t$ region,
we find that $\Upsilon(1S)$ is almost unpolarized at all the $p_t$ range.
The situation for $\Upsilon(2S)$ is similar to $\Upsilon(1S)$, but with more $\chi_{bJ}$ feeddown contribution at small $p_t$ range. 
Therefore, the $\chi_{bJ}$ feeddown contribution is very important to explain the experimental measurement on polarization.
Although the experimental measurement on the fraction of $\chi_{bJ}(1P)$ feeddown in $\Upsilon(1S)$ is already used in the fit, 
it is preliminary with large errors.

It is believed that final physical results are independent of the NRQCD factorization scale $\mu_\Lambda$, 
but the dependence does exist when theoretical calculation is truncated at fix order in the perturbative expansion.
And this dependence can be found when the detailed arrangement of NRQCD factorization formula is taken in the calculation with $P$-wave intermediate state involved.
So a better way to present the final results is to take $\mu_\Lambda$ dependence into the consideration of uncertainty.
In both figures, we have also shown the results with $\mu_\Lambda=m_b$ and
$\mu_\Lambda=\Lambda_{QCD}$.
It is found that the $\mu_\Lambda$ dependence is quite small for the $p_t$ distribution
of $\Upsilon$ yield and polarization due to small contribution of $\Upsilon{}{(\OP{3}{P}{J}{8})}$ and $\chi_{bJ}(\OP{3}{P}{J}{1})$.

In summary, we present the first complete NLO study on the polarization and yield of $\upa$ hadroproduction.
Based on the calculation of the polarization and yield for both direct and feeddown contributions, eleven CO LDMEs are obtained by fitting the experimental data at the Tevatron and LHC step by step for $\Upsilon(3S,2S,1S)$.
With different choices of the NRQCD factorization scale $\mu_\Lambda$, we find that $\mu_\Lambda$ dependence is very small in $p_t$ 
distribution of the yield and polarization for $\Upsilon$ even though it could be quite large for $J/\psi$ where the $P$-wave component contributions are very large.
For $p_t$ distribution of $\Upsilon$ yield, the experimental measurements at the Tevatron and LHC can be explained very well in a wide range of $p_t$.
For $\Upsilon(3S)$, the polarization can not be explained at LO in $v^2$ and NLO in $\alpha_s$ if unknown feeddown contribution 
from higher excited bottomonia is negligible.
For $\Upsilon(1S,2S)$, the predictions for polarization can explain the CMS data well, but still have some distance from the CDF data. 

Further study needs to be considered.
The relativistic corrections to $\jpsi$ hadroproduction~\cite{Xu:2012am} is negative and large in small $p_t$ range, and this infers that the relativistic corrections to $\Upsilon(3S)$ is the largest one among $\upa$ and detailed study may change the result of fit.
The uncertainty from the badly known fraction of $\chi_{bJ}$ feeddown
in the fits for $\Upsilon(1S,2S)$ could be large which is not presented in the plots.
With feeddown contribution of $\chi_{bJ}(3P)$, the polarization of $\Upsilon(3S)$ may be explained as well.  
Therefore a further precise measurement on the fraction of $\chi_{bJ}$ feeddown or on direct $\Upsilon$ production 
will be very helpful to fix the polarization puzzle.

We are thankful for help from the Deepcomp7000 project of the Supercomputing Center, CNIC, CAS and also the TH-1A project of NSCC-TJ. This work is supported, in part, by the National Natural Science Foundation of China (No. 10935012, and No. 11005137),
DFG and NSFC (CRC110), and by CAS under Project No. INFO-115-B01.


\end{document}